\documentclass[aps,twocolumn,prb,showpacs]{revtex4}

\usepackage{graphicx}
\begin{document}
\draft
\preprint{11 March 2007}
\title{Low-energy structure of the intertwining double-chain ferrimagnets\\
       $\mbox{\boldmath$A$}_3$Cu$_3$(PO$_4$)$_4$
       ($\mbox{\boldmath$A$}=\mbox{Ca},\mbox{Sr},\mbox{Pb}$)
       \footnote[2]{Phys. Rev. B {\bf 76}, 014409 (2007)}}
\author{Shoji Yamamoto and Jun Ohara}
\address{Department of Physics, Hokkaido University,
         Sapporo 060-0810, Japan}
%\date{Received \hspace{4cm}}
\date{11 March 2007}
%\date{\today}
\begin{abstract}
Motivated by the homometallic intertwining double-chain ferrimagnets
$A_3$Cu$_3$(PO$_4$)$_4$ ($A=\mbox{Ca},\mbox{Sr},\mbox{Pb}$),
we investigate the low-energy structure of their model Hamiltonian
${\cal H}=\sum_n
 [J_1(\mbox{\boldmath$S$}_{n  :1}+\mbox{\boldmath$S$}_{n  :3})
 +J_2(\mbox{\boldmath$S$}_{n+1:1}+\mbox{\boldmath$S$}_{n-1:3})]
 \cdot\mbox{\boldmath$S$}_{n:2}$,
where $\mbox{\boldmath$S$}_{n:l}$ stands for the $\mbox{Cu}^{2+}$ ion spin
labeled $l$ in the $n$th trimer unit, with particular emphasis on the
range of bond alternation $0<J_2/J_1<1$.
Although the spin-wave theory, whether up to $O(S^1)$ or up to $O(S^0)$,
claims that there exists a flat band in the excitation spectrum regardless
of bond alternation, a perturbational treatment as well as the
exact diagonalization of the Hamiltonian reveals its weak but nonvanishing
momentum dispersion unless $J_2=J_1$ or $J_2=0$.
Quantum Monte Carlo calculations of the static structure factor further
convince us of the low-lying excitation mechanism, elucidating
similarities and differences between the present system and
alternating-spin linear-chain ferrimagnets.
\end{abstract}
% 75.10.Jm: Quantized spin models
% 75.30.Cr: Saturation moments and magnetic susceptibilities 
% 75.30.Ds: Spin waves (for spin-wave resonance, see 76.50.+g)
% 75.40.Cx: Static properties (order parameter, static susceptibility,
%           heat capacities, critical exponents, etc.) 
% 75.40.Gb: Dynamic properties (dynamic susceptibility, spin waves,
%           spin diffusion, dynamic scaling, etc.)
% 75.40.Mg: Numerical simulation studies
% 75.45.$+$j: Macroscopic quantum phenomena in magnetic systems
% 75.50.Gg: Ferrimagnetics
% 75.50.Xx: Molecular magnets
% 76.50.$+$g: Ferromagnetic, antiferromagnetic, and ferrimagnetic
%             resonances; spin-wave resonance
% 76.60.$-$k: Nuclear magnetic resonance and relaxation 
\maketitle

\section{Introduction}

   It is a long-standing and still challenging theme in materials science
to design molecular systems ordering ferromagnetically. \cite{K}
The naivest idea of ferromagnetically coupling nearest-neighbor magnetic
centers leads to the highest spin multiplicity but critically depends on
some structural parameters which are hard to handle chemically.
An alternative solution to highly magnetic ground states consists of
aligning molecular bricks so as to obtain a nonzero resultant spin in the
ground state and then coupling the chains again in a ferromagnetic
fashion.
A variety of quasi-one-dimensional ferrimagnets were thus synthesized and
not a few of them have been attracting theoretical as well as experimental
interest.
\begin{figure}[b]
\vspace*{-4mm}
\centering
\includegraphics[width=76mm]{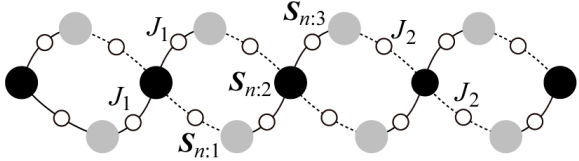}
\vspace*{-3mm}
\caption{Cu$^{2+}$ trimeric chains in $A_3\mbox{Cu}_3(\mbox{PO}_4)_4$.
         The strongly coupled Cu$^{2+}$ trimer consists of a central
         square planar Cu$^{2+}$(1) ion (black circle) and two pyramidal
         Cu$^{2+}$(2) ions (gray circles) bridged by oxygen ions (open
         circles).}
\label{F:illust}
\end{figure}

   Bimetallic chain compounds are early examples and among others is
 $\mbox{MnCu(pbaOH)(H}_2\mbox{O)}_3$
($\mbox{pbaOH}=2\mbox{-hydroxy-}1,3\mbox{-propylenebis(oxamato)}
 =\mbox{C}_7\mbox{H}_6\mbox{N}_2\mbox{O}_7$), \cite{K782}
which retains the long-range ferromagnetic order on the scale of the
crystal lattice.
Replacing the Mn$^{2+}$ ions by Fe$^{2+}$, Co$^{2+}$, and Ni$^{2+}$ ions,
Kahn and co-workers further synthesized a series of isomorphous compounds,
\cite{K3325} which stimulated extensive chemical explorations of
heterometallic chain magnets \cite{K1530,C12837} and systematic
theoretical investigations of alternating-spin chains.
\cite{D413,D10992,K3336,B3921,P8894,Y14008,I14024,M5908,Y1024}
In an attempt to obtain substantially larger couplings between neighboring
magnetic centers and possibly attain transitions to three-dimensional
order at higher temperatures,
Caneschi {\it et al.} \cite{C1976} made a distinct attempt to bring into
interaction metal ions and stable organic radicals.
The representative materials of general formula
$\mbox{Mn(hfac)}_2\mbox{NIT-}R$
($\mbox{hfac}=\mbox{hexafluoroacetylacetonate}
 =\mbox{C}_5\mbox{H}_2\mbox{O}_2\mbox{F}_6$;
 $\mbox{NIT-}R=\mbox{nitronyl\ nitroxide\ radical}
 =\mbox{C}_7\mbox{H}_{12}\mbox{N}_2\mbox{O}_2\mbox{-}R$
 with
 $R=\mbox{CH}_3,\mbox{C}_2\mbox{H}_5,\mbox{C}_3\mbox{H}_5,
    \mbox{C}_6\mbox{H}_5$)
indeed exhibit antiferromagnetic intrachain interactions ranging from
$200$ to $330\,\mbox{cm}^{-1}$.
The metal-radical hybrid strategy, combined with fabrication of novel
polyradicals, \cite{I201} yielded various polymerized heterospin chain
compounds. \cite{O5221,O8067}

   Homometallic ferrimagnetism is also realizable \cite{D3551,A1624} but
its mechanism is often more subtle, essentially depending on the
structural features of the system.
Coronado {\it et al.} \cite{C900,C3907} pioneeringly synthesized
chain-structured compounds of such kind,
$M_2(\mbox{EDTA})(\mbox{H}_2\mbox{O})_4\cdot 2\mbox{H}_2\mbox{O}$
($M=\mbox{Ni},\mbox{Co}$;
 $\mbox{EDTA}=\mbox{ethylenediamminetetraacetate}
 =\mbox{C}_{10}\mbox{N}_2\mbox{O}_8$), whose ferrimagnetic behavior
originates from the alternating $g$ factors and is therefore faint.
Homometallic chain compounds of more pronouncedly ferrimagnetic aspect
\cite{E4466,A5022,C306,R214415} were not obtained until another decade had
passed, where particular topologies were elaborately imposed on the
intrachain exchange interactions.
A series of compounds,
$M(R\mbox{-py})_2(\mbox{N}_3)_2$
($M=\mbox{Cu},\mbox{Mn}$;
 $R\mbox{-py}=\mbox{pyridinic\ ligand}
 =\mbox{C}_5\mbox{H}_4\mbox{N-}R$
 with
 $R=\mbox{Cl},\mbox{CH}_3,\cdots$),
consists of bond-polymerized homometallic chains, where the neighboring
metal ion spins are bridged by versatile azido ligands and are coupled to
each other ferromagnetically or antiferromagnetically.

   The homometallic intertwining double-chain compounds
$A_3\mbox{Cu}_3(\mbox{PO}_4)_4$ ($A=\mbox{Ca},\mbox{Sr},\mbox{Pb}$),
\cite{A29,B395,E6} which are illustrated in Fig. \ref{F:illust}, are
topological ferrimagnets \cite{D83} in the strict sense.
Their hybrid analogs
$\mbox{Ca}_{3-x}\mbox{Sr}_x\mbox{Cu}_3(\mbox{PO}_4)_4$ ($0\leq x\leq 3$)
\cite{B395} were also fabricated in an attempt to tune the
antiferromagnetic bridges between the Cu(1) and Cu(2) sites, labeled $J_1$
and $J_2$, and possibly explore how paramagnetic spins grow into bulk
ferrimagnets.
The magnetic centers without single ion anisotropy and the simple
crystalline structure without any organic ligand will contribute toward
revealing intrinsic features of one-dimensional ferrimagnetic phenomena.
Thus motivated, various experiments have been performed on these copper
phosphates in recent years, including high-field magnetization,
\cite{A186} specific-heat, \cite{B709} inelastic neutron-scattering,
\cite{M144411} nuclear spin-lattice relaxation-time, \cite{Y074703} and
electron-spin-resonance \cite{K094718} measurements.

   It is therefore unfortunate that theoretical investigations of this
system still stay in their early stage. \cite{D83,V185,N214418}
Indeed there exists a field-theoretical study \cite{R4853} deserving
special mention, but the authors restricted their argument to the
particular case of $J_1=J_2$ taking a main interest in realizing organic
ferromagnetism.
A recent numerical diagonalization study \cite{M173} is also a fine guide
to this system, but the authors still devoted themselves to clarifying the
electronic correlation effect on unsaturated ferromagnetism rather than
geometrically modifying this unique bipartite lattice, starting from a
model of the Hubbard type.
An introduction of bond alternation $\delta\equiv J_2/J_1\neq 1$ to this
system will not only contribute toward understanding the magnetic
properties of $A_3\mbox{Cu}_3(\mbox{PO}_4)_4$
\cite{B395,D83,M144411,Y074703} but also illuminate the characteristic of
the uniform point $\delta=1$.
We are thus led to report the whole excitation mechanism of homogeneous-spin
intertwining double-chain ferrimagnets, employing both analytical and
numerical tools.
According to the spin-wave theory, there exist three modes of elementary
excitation, two of which exhibit parallel dispersion relations, while the
rest of which is of no dispersion, regardless of bond alternation.
However, the exact-diagonalization and perturbational calculations
disprove the spin-wave scenario that the low-lying excitation spectrum
remains qualitatively unchanged with varying $\delta$.
Indeed there exist local excitations which are rigorously immobile at
$\delta=1$, but {\it they can be itinerant with $\delta$ moving away from
unity}.
Except for the two particular points $\delta=1$ and $\delta=0$,
corresponding to a plaquette chain and decoupled trimers, respectively,
{\it there is no flat band in the excitation spectrum of the
homogeneous-spin trimeric chain}.
We further inquire into thermal excitations based on such an energy
spectrum.
Calculating the static structure factor as a function of temperature for
an alternating-spin linear-chain ferrimagnet as well as for the present
system, we show what are the universal ferrimagnetic features and how they
vary with decreasing $\delta$.

\section{Plaquette Chains}

   The Hamiltonian of our interest is represented as
\begin{eqnarray}
   &&
   {\cal H}\equiv{\cal H}_1+{\cal H}_2
   =\sum_{n=1}^N
    \bigl[
     J_1(\mbox{\boldmath$S$}_{n:1}+\mbox{\boldmath$S$}_{n:3})
     \cdot\mbox{\boldmath$S$}_{n:2}\ \ \ \ 
   \nonumber\\
   &&\qquad
    +J_2(\mbox{\boldmath$S$}_{n+1:1}+\mbox{\boldmath$S$}_{n-1:3})
     \cdot\mbox{\boldmath$S$}_{n:2}
    \bigr],
   \label{E:H}
\end{eqnarray}
where $\mbox{\boldmath$S$}_{n:l}$ symbolizes the $\mbox{Cu}^{2+}$ ion spin
($S=\frac{1}{2}$) labeled $l$ in the $n$th trimer unit
(see Fig. \ref{F:illust}) and the intratrimer ($J_1$) and intertrimer
($J_2$) exchange interactions, denoted by ${\cal H}_1$ and ${\cal H}_2$,
respectively, are defined as $0\leq J_2\leq J_1$.
First we take a look at the particular point of $J_2/J_1\equiv\delta=1$,
where the model reads a plaquette chain, bearing some analogy with a
linear chain of alternating spins $1$ and $\frac{1}{2}$.
\begin{figure}[b]
\vspace*{-2mm}
\centering
\includegraphics[width=80mm]{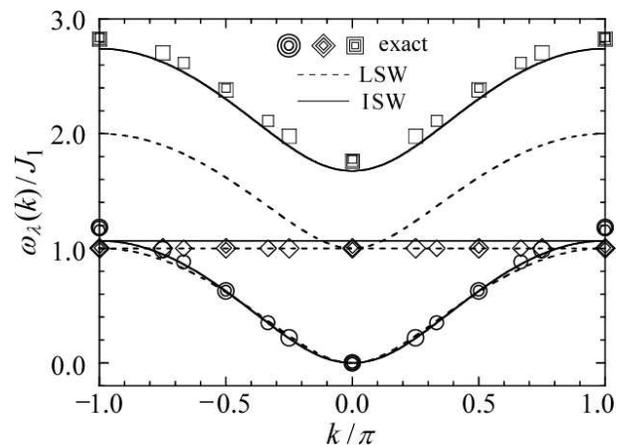}
\vspace*{-3mm}
\caption{Dispersion relations of the elementary excitations in the
         spin-$\frac{1}{2}$ plaquette chain, two (circles and diamonds) of
         which reduce the ground-state magnetization and are thus of
         ferromagnetic character, while the rest (squares) of which
         enhances the ground-state magnetization and is thus of
         antiferromagnetic character.
         The exact-diagonalization results at $N=4$, $6$, and $8$ are
         presented by symbols of small, middle, and large sizes,
         respectively, whereas the up-to-$O(S^1)$ linear (LSW) and
         up-to-$O(S^0)$ interacting (ISW) spin-wave calculations are given
         by dotted and solid lines, respectively.}
\label{F:delta=1}
\end{figure}
\begin{figure*}
\centering
\includegraphics[width=168mm]{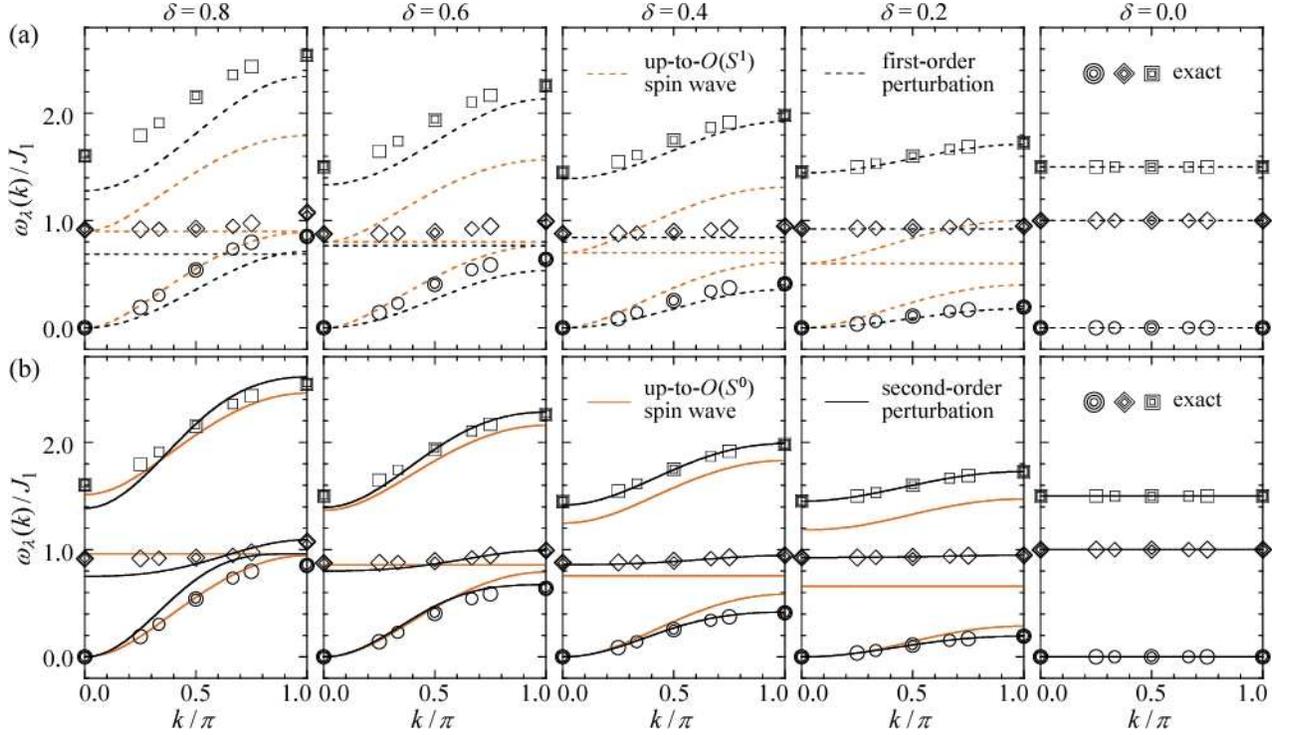}
\vspace*{-3mm}
\caption{(Color online)
         Dispersion relations of the elementary excitations in the
         spin-$\frac{1}{2}$ trimeric chain with varying $\delta$.
         The first-order perturbational calculations, together with the
         up-to-$O(S^1)$ linear spin-wave findings, are given in the upper
         five, whereas the second-order perturbational calculations,
         together with the up-to-$O(S^0)$ interacting spin-wave findings,
         are given in the lower five.
         The exact-diagonalization results at $N=4$, $6$, and $8$ are
         presented in both upper and lower panels by symbols of small,
         middle, and large sizes, respectively}
\label{F:delta<1}
\vspace*{-2mm}
\end{figure*}

   Introducing bosonic operators through the Holstein-Primakoff
transformation
\begin{equation}
   \left.
   \begin{array}{ll}
    \!S_{n:1}^+=\sqrt{2S-a_{n:1}^\dagger a_{n:1}}\ a_{n:1},&
      S_{n:1}^z=S-a_{n:1}^\dagger a_{n:1},\\
    \!S_{n:2}^+=a_{n:2}^\dagger\sqrt{2S-a_{n:2}^\dagger a_{n:2}},&
      S_{n:2}^z=a_{n:2}^\dagger a_{n:2}-S,\\
    \!S_{n:3}^+=\sqrt{2S-a_{n:3}^\dagger a_{n:3}}\ a_{n:3},&
      S_{n:3}^z=S-a_{n:3}^\dagger a_{n:3},\\
   \end{array}
   \right.
   \label{E:HPT}
\end{equation}
defining their Fourier transforms as
\begin{equation}
   a_{k:l}=\frac{1}{\sqrt{N}}\sum_n
           {\rm e}^{{\rm i}(-1)^lk(n+l/2-1)}a_{n:l},
   \label{E:FT}
\end{equation}
with the lattice constant set equal to unity, and further processing
them via the Bogoliubov transformation
\begin{equation}
   \left.
   \begin{array}{lll}
    \alpha_{k:-}^\dagger&=&\psi_{-1}(k)a_{k:1}^\dagger
                        + \psi_{-2}(k)a_{k:2}
                        + \psi_{-3}(k)a_{k:3}^\dagger,\\
    \alpha_{k:0}^\dagger&=&\psi_{01}(k)\,a_{k:1}^\dagger
                        + \psi_{02}(k)\,a_{k:2}
                        + \psi_{03}(k)\,a_{k:3}^\dagger,\\
    \alpha_{k:+}^\dagger&=&\psi_{+1}(k)a_{k:1}
                        + \psi_{+2}(k)a_{k:2}^\dagger
                        + \psi_{+3}(k)a_{k:3},\\
   \end{array}
   \right.
   \label{E:BT}
\end{equation}
we reach a spin-wave Hamiltonian,
\begin{equation}
   {\cal H}
   =E_{\rm g}
   +\sum_{\lambda=\mp,0}
    \omega_\lambda(k)\alpha_{k:\lambda}^\dagger\alpha_{k:\lambda},
\end{equation}
with $E_{\rm g}=\sum_{i=2,1,0}E_{\rm g}^{(i)}$ and
$\omega_\lambda(k)=\sum_{i=1,0}\omega_\lambda^{(i)}(k)$,
where $E_{\rm g}^{(2)}=-2S^2(J_1+J_2)N$ is the classical ground-state
energy, while $E_{\rm g}^{(i)}$ and $\omega_\lambda^{(i)}(k)$
($i=1,0,\cdots$) are the $O(S^i)$ quantum corrections to the ground-state
energy and the dispersion relation of mode $\lambda$, respectively.
Here we have discarded the $O(S^{-1})$ terms.
There are several ways \cite{L5449,I3271,I144429,Y769,C1} of treating the
quartic interactions.
When we diagonalize the one-body terms and then take account of the
two-body terms perturbationally, \cite{Y11033} the spin-wave energies read
\begin{eqnarray}
   &&
   \frac{E_{\rm g}^{(1)}}{J_1N}
   =\frac{S(1+\delta)}{2N}\sum_k\bigl[\omega(k)-3\bigr],
   \label{E:SWEg(1)}
   \\
   &&
   \frac{E_{\rm g}^{(0)}}{J_1N}
   =\frac{1+\delta}{2}({\mit\Gamma}^2-1)
   -\frac{2\delta}{1+\delta}
   \nonumber\\
   &&\qquad\times
    (3{\mit\Gamma}^2+2{\mit\Lambda}^2
    -5{\mit\Gamma}{\mit\Lambda}
    -3{\mit\Gamma}+3{\mit\Lambda}),
   \label{E:SWEg(0)}
   \\
   &&
   \frac{\omega_\mp^{(1)}(k)}{J_1}
   =\frac{S(1+\delta)}{2}\bigl[\omega(k)\mp 1\bigr],
   \nonumber\\
   &&
   \frac{\omega_0^{(1)}(k)}{J_1}=S(1+\delta),
   \label{E:SWdsp(1)}
   \\
   &&
   \frac{\omega_\mp^{(0)}(k)}{J_1}
   =\frac{1+\delta}{2}
    \bigl[{\mit\Gamma}{\mit\Gamma}(k)\mp{\mit\Gamma}\bigr]
   -\frac{\delta}{1+\delta}
   \nonumber\\
   &&\qquad\times
   \bigl[
    6{\mit\Gamma}{\mit\Gamma}(k)
   -5{\mit\Gamma}{\mit\Lambda}(k)
   -5{\mit\Lambda}{\mit\Gamma}(k)
   +4{\mit\Lambda}{\mit\Lambda}(k)\ \ \ \ 
   \nonumber\\
   &&\qquad\ \,
   -3{\mit\Gamma}(k)+3{\mit\Lambda}(k)
   \mp{\mit\Gamma}\pm{\mit\Lambda}
   \bigr],
   \nonumber\\
   &&
   \frac{\omega_0^{(0)}(k)}{J_1}
   =-\frac{1+\delta}{2}({\mit\Gamma}-1)
    -\frac{2\delta}{1+\delta}({\mit\Gamma}-{\mit\Lambda}),
   \label{E:SWdsp(0)}
\end{eqnarray}
and their eigenvectors are given by
\begin{eqnarray}
   &&
   \psi_{\mp 1}(k)=\psi_{\mp 3}^*(k)
   =\frac{2({\rm e}^{\pm{\rm i}k/2}+\delta{\rm e}^{\mp{\rm i}k/2})}
         {(1+\delta)\sqrt{2\omega(k)\bigl[3\mp\omega(k)\bigr]}},\ \ \ \ 
   \nonumber\\
   &&
   \psi_{\mp 2}(k)
   =\sqrt{\frac{3\mp\omega(k)}{2\omega(k)}},\ \ 
   \psi_{01}(k)=\frac{1}{\sqrt{2}},\ \ 
   \nonumber\\
   &&
   \psi_{02}(k)=0,\ \ 
   \psi_{03}(k)
   =-\frac{{\rm e}^{-{\rm i}k/2}+\delta{\rm e}^{{\rm i}k/2}}
          {\sqrt{2}({\rm e}^{{\rm i}k/2}+\delta{\rm e}^{-{\rm i}k/2})},
   \label{E:psi}
\end{eqnarray}
where
\begin{eqnarray}
   &&
   \omega(k)
   =\sqrt{1+\frac{32\delta}{(1+\delta)^2}\sin^2\frac{k}{2}},
   \label{E:omega(k)}
   \\
   &&
   {\mit\Gamma}
   =\frac{1}{N}\sum_k{\mit\Gamma}(k)
   =\frac{1}{N}\sum_k\frac{1}{\omega(k)},
   \nonumber\\
   &&
   {\mit\Lambda}
   =\frac{1}{N}\sum_k{\mit\Lambda}(k)
   =\frac{1}{N}\sum_k\frac{\cos k}{\omega(k)}.
   \label{E:Gamma}
\end{eqnarray}

   Figure \ref{F:delta=1} shows the thus-calculated spin-wave excitation
modes together with the exact eigenvalues.
Free spin waves well describe the ferromagnetic modes
$\omega_-(k)$ and $\omega_0(k)$, while higher-order quantum corrections
play an essential role in reproducing the antiferromagnetic mode
$\omega_+(k)$.
The $O(S^0)$ quantum corrections significantly improve fully delocalized
magnetic excitations in general, \cite{O8067,Y13610,I3271} but the
standard Holstein-Primakoff magnon series expansion seems not to work well
for highly localized excitations.
The dispersive branches $\omega_\mp(k)$ are nothing but the
elementary excitation modes of spin-alternating linear-chain Heisenberg
ferrimagnets. \cite{Y211}
They are parallel within the spin-wave theory, but their difference
$\omega_+(k)-\omega_-(k)$ is momentum dependent in fact.
On the other hand, the flat band $\omega_0(k)$ arises from further
excitation degrees of freedom in the present system.
When $J_1=J_2$, the Hamiltonian (\ref{E:H}) reads
\begin{equation}
   {\cal H}
   =J_1\sum_{n=1}^{N}
   (\mbox{\boldmath$S$}_{n:2}+\mbox{\boldmath$S$}_{n+1:2})
   \cdot\mbox{\boldmath$T$}_{n:3;n+1:1},
   \label{E:Hdelta=1}
\end{equation}
with composite spins
$\mbox{\boldmath$T$}_{n:3;n+1:1}\equiv
 \mbox{\boldmath$S$}_{n:3}+\mbox{\boldmath$S$}_{n+1:1}$,
each lying diagonally across an elementary palquette.
Since the Hamiltonian (\ref{E:Hdelta=1}) commutes with
$\mbox{\boldmath$T$}_{n:3;n+1:1}^2\equiv T_{n:3;n+1:1}(T_{n:3;n+1:1}+1)$,
we have good quantum numbers $T_{n:3;n+1:1}$, each taking either 0 or 1.
Therefore, the plaquette-chain Hamiltonian is block-diagonalized by 
the set of numbers $\{T_{n:3;n+1:1};\,n=1,2,\cdots,N\}$ \cite{T6405}
as well as by the total magnetization
$\sum_{n=1}^N(S_{n:2}^z+T_{n:3;n+1:1}^z)\equiv{\cal M}$.
The Hilbert space of
$\sum_{n=1}^N(\mbox{\boldmath$T$}_{n:3;n+1:1})^2/2
=\sum_{n=1}^N(\mbox{\boldmath$S$}_{n:3}\cdot
              \mbox{\boldmath$S$}_{n+1:1}+3/4)
\equiv{\cal N}=N$
corresponds to the ferrimagnetic chain of alternating spins $1$ and
$\frac{1}{2}$ and consequently we here have exactly the same dispersion
relations \cite{Y13610} of elementary excitations.
The Hilbert space of ${\cal N}=N-1$ and ${\cal M}=N/2-1$ consists of $N$
subspaces labeled
$\{T_{1:3;2:1},T_{2:3;3:1},\cdots,T_{N:3;1:1}\}
=\{0,1,\cdots,1\},\{1,0,1,\cdots,1\},\cdots,\{1,\cdots,1,0\}$, and they
all give the same set of eigenvalues, forming $N$ flat bands.
We find the lowest one in Fig. \ref{F:delta=1}.
\begin{figure}[b]
\centering
\includegraphics[width=76mm]{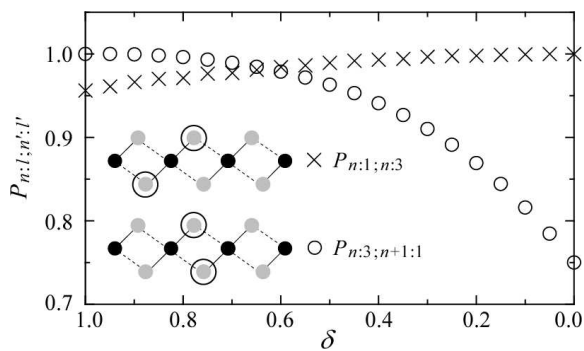}
\vspace*{-3mm}
\caption{Probability of two spin $\frac{1}{2}$'s constructing a spin $1$
         in the ground state of the spin-$\frac{1}{2}$ trimeric chain of
         $N=64$ with varying $\delta$, where
         $P_{n:3;n+1:1}\equiv\mbox{\boldmath$T$}_{n:3;n+1:1}^2/2=
          \mbox{\boldmath$S$}_{n:3}\cdot\mbox{\boldmath$S$}_{n+1:1}+3/4$
         and
         $P_{n:1;n:3}\equiv\mbox{\boldmath$T$}_{n:1;n:3}^2/2=
          \mbox{\boldmath$S$}_{n:1}\cdot\mbox{\boldmath$S$}_{n:3}+3/4$
         are estimated by a quantum Monte Carlo method.}
\label{F:projection}
\vspace*{-2mm}
\end{figure}

   Thus and Thus, the spin-$S$ plaquette chain turns out a combination of
the alternating-spin-$(2S,S)$ linear chain and extra excitation degrees of
freedom within the composite spins $\mbox{\boldmath$T$}_{n:3;n+1:1}$.
All the composite spins are saturated in the ground state,
$\mbox{\boldmath$T$}_{n:3;n+1:1}^2=2S(2S+1)$, and therefore, their
excitations are necessarily of ferromagnetic aspect.
The ferromagnetic excitations of local character are well understandable
within the spin-wave description.
Equations (\ref{E:BT}) and (\ref{E:psi}) show that
$a_{n:1}^\dagger$ and $a_{n:3}^\dagger$, creating bosonic excitations on
the Cu$^{2+}$(2) sites, indeed participate in the construction of
$\alpha_{k:0}^\dagger$, but any of $a_{n:2}^\dagger$, creating bosonic
excitations on the Cu$^{2+}$(1) sites, does not.
Without mediation of bridging spins $\mbox{\boldmath$S$}_{n:2}$, any
intraplaquette excitation is never movable.
Then what may happen with $\delta$ moving away from unity?
The spin-wave theory, whether up to $O(S^1)$ or up to $O(S^0)$, predicts
that the ferromagnetic and antiferromagnetic excitation modes
$\omega_\mp(k)$ are still parallel and the extra ferromagnetic excitation
mode between them, $\omega_0(k)$, remains dispersionless.
Let us verify the true scenario.

\section{Bond-Alternating Trimeric Chains}

   We demonstrate in Fig. \ref{F:delta<1} several schemes of calculating
low-lying excitation modes for the spin-$\frac{1}{2}$ bond-alternating
trimeric chain.
In spite of the persistent flat band within the spin-wave theory,
the exact diagonalization reveals that it can be dispersive with varying
$\delta$.
When $\delta\neq 1$, the Hamiltonian (\ref{E:H}) does not commute with
$\mbox{\boldmath$T$}_{n:3;n+1:1}^2$.
Now that there is a certain probability of composite spins
$\mbox{\boldmath$T$}_{n:3;n+1:1}$ being singlet even in the ground state,
the gapped ferromagnetic excitation mode $\omega_0(k)$ is no more
describable as their individual triplet-to-singlet flips.
At $\delta=0$, any excitation is localized within a trimer of
$\mbox{\boldmath$S$}_{n:1}$, $\mbox{\boldmath$S$}_{n:2}$, and
$\mbox{\boldmath$S$}_{n:3}$, and the excitation spectrum degenerates into
three flat bands, $\omega_-(k)\equiv 0$, $\omega_0(k)=J_1$, and
$\omega_+(k)=3J_1/2$.
Figure \ref{F:delta<1} shows that the middle branch of them connects with
the flat band at $\delta=1$.
Without $J_2$, the Hamiltonian (\ref{E:H}) is reduced to
\begin{equation}
   {\cal H}={\cal H}_1
   =J_1\sum_{n=1}^{N}
    \mbox{\boldmath$S$}_{n:2}\cdot\mbox{\boldmath$T$}_{n:1;n:3},
   \label{E:Hdelta=0}
\end{equation}
with intratrimer composite spins
$\mbox{\boldmath$T$}_{n:1;n:3}\equiv
 \mbox{\boldmath$S$}_{n:1}+\mbox{\boldmath$S$}_{n:3}$
and thus commutes with $\mbox{\boldmath$T$}_{n:1;n:3}^2$.
The plaquette chain (\ref{E:Hdelta=1}) and the decoupled trimers
(\ref{E:Hdelta=0}) both exhibit a flat band due to gapped ferromagnetic
excitations, but their ways of constructing local immobile excitations are
different from each other.
Triplet-to-singlet [$(4S+1)$-fold-multiplet-breaking in general] flips of
intraplaquette composite spins $\mbox{\boldmath$T$}_{n:3;n+1:1}$ are
the elementary excitations in the former, while those of intratrimer
composite spins $\mbox{\boldmath$T$}_{n:1;n:3}$ are the elementary
excitations in the latter.
Figure \ref{F:projection} shows how such composite spins behave in the
ground state with varying $\delta$.
Neither $\mbox{\boldmath$T$}_{n:3;n+1:1}$
nor $\mbox{\boldmath$T$}_{n:1;n:3}$ form complete triplets at
$0<\delta<1$, due to nonvanishing off-diagonal matrix elements
$\langle T_{n:3;n+1:1}=1|{\cal H}|T_{n:3;n+1:1}=0\rangle$ and
$\langle T_{n:1;n:3}=1|{\cal H}|T_{n:1;n:3}=0\rangle$.
At the two particular points $\delta=1$ and $\delta=0$, only the
Cu$^{2+}$(2) ion spins
$\mbox{\boldmath$S$}_{n:1}$ and $\mbox{\boldmath$S$}_{n:3}$
constitute the gapped ferromagnetic excitation mode, but otherwise the
Cu$^{2+}$(1) ion spins $\mbox{\boldmath$S$}_{n:2}$ also contribute to
that.
Without interconnecting spins $\mbox{\boldmath$S$}_{n:2}$,
any excitation is immobile, whereas with their mediation, all the local
excitations can be itinerant and the resultant bands are dispersive.

   Perturbational calculations support such a scenario.
With increasing couplings $J_2$ between isolated trimers, the energy
dispersion relations grow as follows:
\begin{eqnarray}
   &&
   \frac{E_{\rm g}}{J_1N}
   =-1-\frac{\delta}{9}-\frac{869}{2430}\delta^2+O(\delta^3),
   \label{E:pertEg}
   \\
   &&
   \frac{\omega_-(k)}{J_1}
   =\frac{4}{9}\delta(1-\cos k)+\frac{\delta^2}{2430}(929-474
   \nonumber\\
   &&\qquad\times
   \cos k-455\cos 2k)+O(\delta^3),
   \label{E:pertdsp-}
   \\
   &&
   \frac{\omega_0(k)}{J_1}
   =1-0.38970\delta+\delta^2(0.32099-0.26736
   \nonumber\\
   &&\qquad\times
   \cos k+0.04212\cos 2k)
   +O(\delta^3),
   \label{E:pertdsp0}
   \\
   &&
   \frac{\omega_+(k)}{J_1}
   =\frac{3}{2}+\frac{\delta}{18}(7-12\cos k)
   +\frac{\delta^2}{810}(346-100\ \ \ \ 
   \nonumber\\
   &&\qquad\times
   \cos k-109\cos 2k)+O(\delta^3).
   \label{E:pertdsp+}
\end{eqnarray}
Equations (\ref{E:pertdsp-})-(\ref{E:pertdsp+}) are also drawn in
Fig. \ref{F:delta<1}.
The first-order perturbation points out that not only $\omega_\mp(k)$
themselves but also their difference should be dispersive, but it cannot
reveal nonvanishing momentum dependence of $\omega_0(k)$.
We cannot reproduce the dispersive middle band until we take account of
the second-order perturbation.
%This is because the gapped ferromagnetic excitation mode of decoupled
%trimers is heavily degenerate.
The lowest ferromagnetic and antiferromagnetic excitations of decoupled
trimers (\ref{E:Hdelta=0}), gapless and gapped by $3J_1/2$ from the ground
state, respectively, are both $N$-fold degenerate and are expressed as
\begin{eqnarray}
   &&
   |E_\mp(m)\rangle
   =|-(1\pm 3)J_1/4;1/2\mp 1\rangle_m
   \nonumber\\
   &&\qquad
   \mathop{\otimes}_{n\neq m}|-J_1;1/2\rangle_n
   \ \ (m=1,2,\cdots N),\ \ \ \ 
\end{eqnarray}
while their gapped ferromagnetic excitations at an energy cost of
$J_1$ are $N^2$-fold degenerate and are expressed as
\begin{eqnarray}
   &&
   |E_0(m,m')\rangle
   =\delta_{mm'}
    |0;-1/2\rangle_m\mathop{\otimes}_{n\neq m}|-J_1;1/2\rangle_n
   \nonumber\\
   &&\qquad
   +(1-\delta_{mm'})
    |0;1/2\rangle_m\otimes|-J_1;-1/2\rangle_{m'}
   \nonumber\\
   &&\qquad
   \mathop{\otimes}_{n\neq m,m'}|-J_1;1/2\rangle_n
   \ \ (m,m'=1,2,\cdots N),\ \ \ \ 
\end{eqnarray}
in terms of the eigenstates of an isolated trimer
$|\mbox{\boldmath$S$}_{n:1},
  \mbox{\boldmath$S$}_{n:2},
  \mbox{\boldmath$S$}_{n:3}\rangle$,
\begin{eqnarray}
   &&
   |-J_1;1/2\rangle_n
   =\frac{1}{\sqrt{6}}
    (|\!\uparrow\uparrow\downarrow\rangle
    -2|\!\uparrow\downarrow\uparrow\rangle
    +|\!\downarrow\uparrow\uparrow\rangle),
   \nonumber\\
   &&
   |-J_1;-1/2\rangle_n
   =\frac{1}{\sqrt{6}}
    \bigl(
     |\!\downarrow\downarrow\uparrow\rangle
    -2|\!\downarrow\uparrow\downarrow\rangle
    +|\!\uparrow\downarrow\downarrow\rangle
    \bigr),
   \nonumber\\
   &&
   |0;1/2\rangle_n
   =\frac{1}{\sqrt{2}}
    \bigl(
     |\!\uparrow\uparrow\downarrow\rangle
    -|\!\downarrow\uparrow\uparrow\rangle
    \bigr),
   \nonumber\\
   &&
   |0;-1/2\rangle_n
   =\frac{1}{\sqrt{2}}
    \bigl(
     |\!\downarrow\downarrow\uparrow\rangle
    -|\!\uparrow\downarrow\downarrow\rangle
    \bigr),
   \nonumber\\
   &&
   |J_1/2;3/2\rangle_n
   = |\!\uparrow\uparrow\uparrow\rangle,
   \nonumber\\
   &&
   |J_1/2;1/2\rangle_n
   =\frac{1}{\sqrt{3}}
    \bigl(
     |\!\uparrow\uparrow\downarrow\rangle
    +|\!\uparrow\downarrow\uparrow\rangle
    +|\!\downarrow\uparrow\uparrow\rangle
    \bigr),
   \nonumber\\
   &&
   |J_1/2;-1/2\rangle_n
   =\frac{1}{\sqrt{3}}
    \bigl(
     |\!\uparrow\downarrow\downarrow\rangle
    +|\!\downarrow\uparrow\downarrow\rangle
    +|\!\downarrow\downarrow\uparrow\rangle
    \bigr),
   \nonumber\\
   &&
   |J_1/2;-3/2\rangle_n
   = |\!\downarrow\downarrow\downarrow\rangle.
\end{eqnarray}
With perturbational interactions ${\cal H}_2$ turned on,
the $N$-fold degeneracy of the eigenvalue
$-[N-3(1\mp 1)/4]J_1=\langle E_\mp(m)|{\cal H}_1|E_\mp(m)\rangle$
is completely lifted, whereas the $N^2$-fold degenerate eigenvalue
$-(N-1)J_1=\langle E_0(m,m')|{\cal H}_1|E_0(m,m')\rangle$
only splits into $N$ flat bands within the first-order corrections.
The second-order corrections are necessary for reproducing the dispersion
relation of $\omega_0(k)$.
In this context we may be reminded that Honecker and L\"auchli
\cite{H174407} pioneeringly investigated analogous but frustrated
Cu$^{2+}$ trimeric chains.
The gapless ferromagnetic excitation mode (\ref{E:pertdsp-}) is indeed
derived from their effective Hamiltonian under strong trimerization
$\delta\ll 1$.

\section{Summary and Discussion}

   We have investigated the low-energy structure of intertwining
double-chain ferrimagnets composed of homogeneous spins with particular
emphasis on the gapped ferromagnetic excitation mode.
While there exist a macroscopic number of flat bands \cite{M1851} in the
excitation spectrum at $\delta=1$ and $\delta=0$, which signify
uncorrelated excitations of local spin-$2S$ multiplets in any case,
they become dispersive as soon as $\delta$ moves away from these
particular points.
Pair excitations of corner spins
$\mbox{\boldmath$S$}_{n:3}$ and $\mbox{\boldmath$S$}_{n+1:1}$
are elementary in plaquette chains of $\delta=1$, while those of
$\mbox{\boldmath$S$}_{n:1}$ and $\mbox{\boldmath$S$}_{n:3}$
are elementary in decoupled trimers of $\delta=0$, both of which are
completely immobile without any mediation of joint spins
$\mbox{\boldmath$S$}_{n:2}$.
The spin-wave theory successfully characterizes the plaquette chain but
fails to find arising contribution of $\mbox{\boldmath$S$}_{n:2}$ to
gapped ferromagnetic excitations with bond alternation.
Such a misleading prediction has been corrected by further numerical and
analytical investigations.
\begin{figure*}
\centering
\includegraphics[width=168mm]{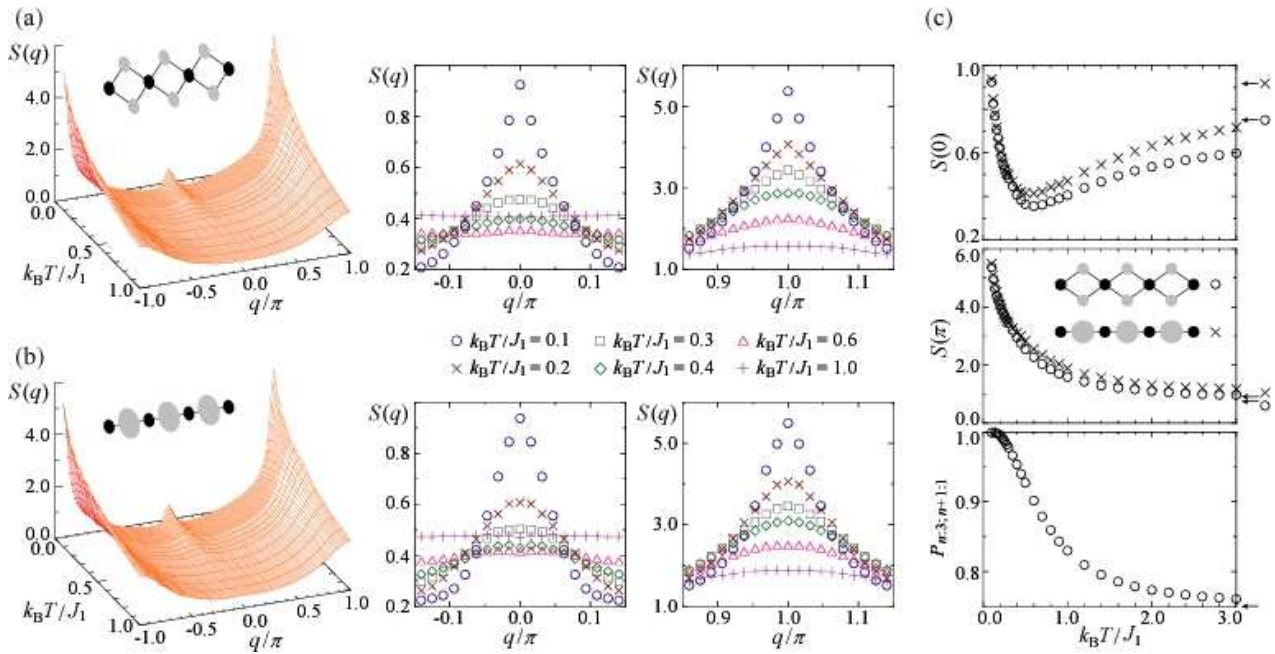}
\vspace*{-3mm}
\caption{(Color online)
         Quantum Monte Carlo calculations of the static structure factor
         $S(q)$, with the distance between neighboring spins in the chain
         direction set equal to unity, as a function of temperature.
         The whole view and enlargements at $q=0$ and $q=\pi$ for
         the spin-$\frac{1}{2}$ plaquette chain of $N=64$ (a) and the
         alternating-spin-$(1,\frac{1}{2})$ linear chain of $N=64$ (b).
         The ferromagnetic [$S(0)$] and antiferromagnetic [$S(\pi)$]
         peaks are observed in more detail (c), where the common
         asymptotic values in the high-temperature limit, $3/4$ and
         $11/12$ for the the spin-$\frac{1}{2}$ plaquette chain and the
         alternating-spin-$(1,\frac{1}{2})$ linear chain, respectively,
         are indicated with arrows.
         Thermal averages of the projection
         $P_{n:3;n+1:1}\equiv\mbox{\boldmath$T$}_{n:3;n+1:1}^2/2=
          \mbox{\boldmath$S$}_{n:3}\cdot\mbox{\boldmath$S$}_{n+1:1}+3/4$
         in the spin-$\frac{1}{2}$ plaquette chain are also shown for
         reference, where the asymptotic value in the high-temperature
         limit, $3/4$, is indicated with an arrow.}
\label{F:S(q)delta=1}
\vspace*{-2mm}
\end{figure*}
\begin{figure}[b]
\centering
\includegraphics[width=82mm]{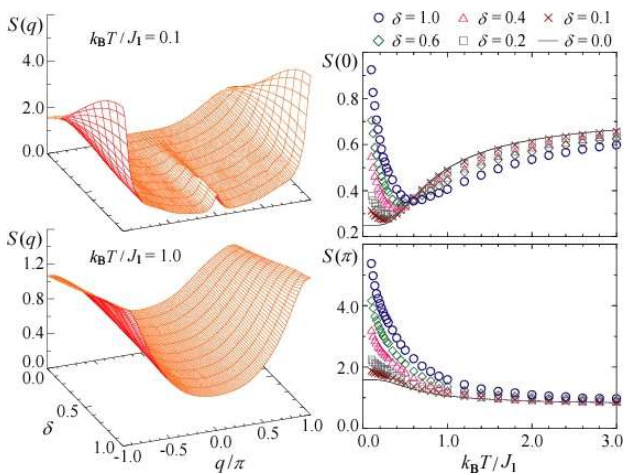}
\vspace*{-3mm}
\caption{(Color online)
         Quantum Monte Carlo calculations of the static structure factor
         $S(q)$, with the distance between neighboring spins in the chain
         direction set equal to unity, as a function of bond alternation
         and temperature for the spin-$\frac{1}{2}$ trimeric chain of
         $N=64$.}
\label{F:S(q)delta<1}
\end{figure}

   The spin-$S$ plaquette chain thus shares whole the nature of the
alternating-spin-$(2S,S)$ linear chain and further exhibits ferromagnetic
excitations of its own.
All the findings but the flat band in Fig. \ref{F:delta=1} are indeed
exactly the same as we have in the ferrimagnetic Heisenberg chain of
alternating spins $1$ and $\frac{1}{2}$. \cite{Y11033}
Even though the homogeneous-spin plaquette chain and the alternating-spin
linear chain are equivalent in their ground states, the former
demonstrates its extra excitation degrees of freedom and deviates from the
latter with increasing temperature.
In order to illuminate similarities and differences between them, we show
in Fig. \ref{F:S(q)delta=1} quantum Monte Carlo calculations of their
static structure factors
\begin{equation}
   S(q)=\frac{1}{N}\sum_{n,l,n',l'}
        {\rm e}^{{\rm i}q(x_{n:l}-x_{n':l'})}S_{n:l}^zS_{n':l'}^z,
\end{equation}
as functions of temperature, where the chain-directional coordinates
$x_{n:l}$ are given in the unit of neighboring-spin spacing.
The pronounced peaks at $q=0$ and $q=\pi$ reflect the ferromagnetic and
antiferromagnetic double excitation mechanism in common.
Without any field applied, $S(0)$ and $S(\pi)$ are, respectively, the
uniform and the staggered susceptibilities multiplied by temperature.
With decreasing temperature, they both diverge as $1/T$.
\cite{Y11033,R4853,Y054423}
With increasing temperature, they both approach the paramagnetic value
$\sum_lS_{n:l}(S_{n:l}+1)/3$ but behave differently at intermediate
temperatures.
A minimum of $S(0)$ as a function of temperature is characteristic of
ferrimagnets. \cite{D413,D10992,O8067,C306,D83,Y064426,G134427,F104454}
$S(0)$ monotonically decreases and increases with increasing temperature
in ferromagnets and antiferromagnets, respectively. \cite{Y1024}
Though the thermal as well as quantum behaviors of the spin-$S$ plaquette
chain and the alternating-spin-$(2S,S)$ linear chain are very much alike,
yet there grows a difference between them with pair excitations of
intraplaquette spins
$\mbox{\boldmath$S$}_{n:3}$ and $\mbox{\boldmath$S$}_{n+1:1}$ from their
highest multiplets.
The ferromagnetic and antiferromagnetic structures of $S(q)$ less survive
increasing temperature in the spin-$S$ plaquette chain than in the
alternating-spin-$(2S,S)$ linear chain.
The larger $S$ the major difference in $S(q)$ as
${\rm lim}_{T\rightarrow\infty}[S^{(2S,S)}(q)-S^{(S,S,S)}(q)]=2S^2/3$.
Alternating-spin-$(2S,S)$ ferrimagnetic chains behave like combinations of
spin-$S$ ferromagnetic and spin-$(2S)$ antiferromagnetic chains,
\cite{Y1024} while such a simple magnetic sum rule is not available to
intertwining double-chain ferrimagnets of our interest.
Additional intraplaquette antiferromagnetic interactions induce
incommensurate peaks in $S(q)$, \cite{G} making corner spins
$\mbox{\boldmath$S$}_{n:3}$ and $\mbox{\boldmath$S$}_{n+1:1}$ frustrated.

   Once $\delta$ moves away from unity, the homogeneous-spin trimeric
chain never more shares any feature of the alternating-spin chain.
Figure \ref{F:S(q)delta<1} presents $S(q)$ with varying $\delta$ and
analyzes its features at $q=0$ and $q=\pi$ in particular.
At low temperatures, $S(0)$ and $S(\pi)$ both decline with decreasing
$\delta$, but they still diverge as $1/T$ unless $\delta=0$. \cite{D83}
At high temperatures, $S(\pi)$ remains decreasing, whereas $S(0)$ turns
increasing, with decreasing $\delta$.
Decoupled trimers are nothing more than paramagnets and their structure
factor is given as
\begin{eqnarray}
   &&
   S(q)=\frac{3}{4}
       -\frac{2}{3}
        \frac{{\rm e}^{J_1/k_{\rm B}T}-{\rm e}^{-J_1/2k_{\rm B}T}}
             {{\rm e}^{J_1/k_{\rm B}T}+1+2{\rm e}^{-J_1/2k_{\rm B}T}}
        \cos q\ \ \ \ 
   \nonumber\\
   &&\qquad
       +\frac{1}{6}
        \frac{{\rm e}^{J_1/k_{\rm B}T}-3+2{\rm e}^{-J_1/2k_{\rm B}T}}
             {{\rm e}^{J_1/k_{\rm B}T}+1+2{\rm e}^{-J_1/2k_{\rm B}T}}
        \cos 2q,
   \label{E:S(q)delta=0}
\end{eqnarray}
which is also drawn in Fig. \ref{F:S(q)delta<1} with solid lines.
Equation (\ref{E:S(q)delta=0}) at $q=0$ reads as the effective Curie law
for a trimer entity, where the Curie constant varies from $1/4$,
attributable to the ground-state doublet, to $3/4$, simply coming from
free spin $\frac{1}{2}$'s, with increasing temperature.
Arising intertrimer couplings $J_2$ immediately pronounce a quadratic
dispersion relation of the ferromagnetic excitations at small momenta
and their further increase costs the antiferromagnetic excitations
higher energy.
That is why growing global correlations enhance and reduce the uniform
susceptibility-temperature product at low and high temperatures,
respectively.

   Weak but nonvanishing dispersion of the gapped ferromagnetic excitation
mode is the most remarkable findings of ours and is the very
characteristic of intertwining double-chain ferrimagnets.
As the existent compounds $A_3$Cu$_3$(PO$_4$)$_4$ have all been reported
to exhibit rather strong bond alternation $\delta\alt 0.1$,
\cite{D83,B709,M144411,Y074703,K094718} it may be hard to detect the
dispersion relation $\omega_0(k)$ there.
Ni analogs, if available, will present an energy structure of the same
type on an enlarged energy scale.
Highly localized excitations in fully exchange-coupled bulk magnets may
either arise from an accidental arrangement of exchange couplings or come
out of a particular lattice structure of geometric aspect.
While some examples of the former case can be found in spin-$\frac{1}{2}$
bond-polymerized chains in principle, \cite{N214418,H014414} it must be
hard to observe them in real materials.
On the other hand, the present findings are of the latter origin and may
be more accessible experimentally.
The Shastry-Sutherland lattice \cite{S1069} is also interesting in this
sense and the model compound SrCu$_2$(BO$_3$)$_2$ indeed exhibits an
excitation mode of little dispersion in its low-energy spectrum.
\cite{K5876}
Such local excitations crystallize to form a superlattice, quantizing the
ground-state magnetization. \cite{M3231,M3417,O1016}
There may be a similar scenario in low-dimensional ferrimagnets of
topological origin as well.
We hope the present study will stimulate further experimental
explorations.
\vspace*{4mm}

\acknowledgments

\vspace*{-4mm}
   The authors are grateful to T. Hikihara for valuable comments.
This work was supported by the Ministry of Education, Culture, Sports,
Science, and Technology of Japan.

\end{document}